\documentclass[aps,pre,twocolumn,showpacs,superscriptaddress]{revtex4-2}

\usepackage[dvipsnames,svgnames,x11names,table]{xcolor}

\usepackage{hyperref}
\usepackage{amsmath}
\usepackage{amssymb}
\usepackage{mathtools}
\usepackage{graphicx}

\usepackage[arrow, matrix, curve]{xy}
\usepackage{bm}
\usepackage{yhmath}
\usepackage{url}
\usepackage{here}
\usepackage{comment}
\newcommand\diff{\mathrm{d}}

\hypersetup{colorlinks=true, linkcolor=BrickRed, urlcolor=blue!50!black, citecolor=blue!50!black}

\usepackage[normalem]{ulem}

\makeatletter
\newcommand\hide@visible[1]{%
  \bgroup\fboxsep=.3ex\colorbox{Gray}{begin hide}%
  #1\colorbox{Gray}{end hide}\egroup%
}
\newcommand\hide@hidden[1]{%
  \bgroup\fboxsep=.3ex\colorbox{Gray}{hidden text}%
}
\newcommand\hide@invisible[1]{}
\newcommand\makevisible{\let\hide\hide@visible}
\newcommand\makehidden{\let\hide\hide@hidden}
\newcommand\makeinvisible{\let\hide\hide@invisible}
\makeatother
\makehidden

\usepackage[capitalise,nameinlink]{cleveref}

\crefname{section}{Sec.}{Secs.} 
\usepackage{graphicx}
\usepackage[caption=false]{subfig} 

\usepackage{verbatim} 
\usepackage{natbib}
\usepackage{bbm}

\begin{document}

 \title{Intermediate scattering function of Brownian particles in a tilted cosine potential}
\author{Regina Rusch}
\affiliation{Institut f\"ur Theoretische Physik, Technikerstra{\ss}e 21-A, Universit\"at Innsbruck, A-6020 Innsbruck, Austria}
\author{Thomas Franosch}
\affiliation{Institut f\"ur Theoretische Physik, Technikerstra{\ss}e 21-A, Universit\"at Innsbruck, A-6020 Innsbruck, Austria}

\date{\today}

\begin{abstract}
We solve the Fokker-Planck equation for a Brownian particle in a tilted cosine potential and derive the intermediate scattering function (ISF), which captures the full spatio-temporal dynamics of the system. The model consists of a single overdamped Brownian particle in one dimension. We derive a generalized ISF comprising two wave vectors to describe correlations in the periodic potential. Exploiting the periodicity via Bloch's theorem, we formulate the problem within a spectral-theoretical framework and numerically compute the corresponding eigenfunctions and eigenvalues, from which we obtain the ISF and the probability density. Using time-dependent perturbation theory, we expand the ISF and derive low-order moments, including the mean-square displacement, time-dependent diffusivity, skewness, and the non-Gaussian parameter. Our analytical results are validated by Brownian-dynamics simulations and analyzed focussing on different regimes of the tilting force. The results are compared to a harmonic approximation and the deterministic limit.

\end{abstract}

\maketitle
\section{Introduction}

Brownian particles in periodic structures exhibit a range of transport phenomena and nontrivial dynamics that are both of fundamental interest in biological and statistical physics and of technological relevance~\cite{Hanggi_2009,Ghosh_2013}. Their dynamics have been studied in a variety of physical systems, including pendula, Josephson junctions, superionic conductors, dipoles in constant fields, and phase-locked loops~\cite{ladera2025tilted}, as reviewed in detail by Risken~\cite{risken1996fokker}. Moreover, these systems provide simple yet powerful models for biological molecular motors~\cite{ait2003brownian}, ion channels~\cite{chou1999entropy}, and transport in porous media~\cite{ledesma2016generalized}.

A Brownian particle in a tilted cosine potential, also known as tilted washboard potential, represents a simple, non-equilibrium system, exhibiting drift, diffusion, and barrier hopping under the influence of thermal noise. Previous works have studied the mean drift velocity~\cite{CHENG_2015}, the effective diffusion constant~\cite{Reimann_2001,Berezhkovskii_2019} or have analyzed the band structure of the corresponding Fokker-Planck operator~\cite{Alamilla_2020}. 
A particularly well-studied effect is giant diffusion, in which the long-time diffusion coefficient is strongly enhanced relative to the bare diffusion constant~\cite{Iida_2025,Reimann_2001,Reimann_2002}.

Another linked phenomenon is gravitaxis, which describes the directed motion of active particles or microorganisms in response to gravity and is particularly relevant in biological systems, such as algae and bacteria~\cite{ten2014gravitaxis,Kuemmel_2013}. In this context, the angular dynamics have been modeled using tilted periodic potentials, with gravity acting as the external tilt~\cite{Chepizhko_2022,Rusch_2024}.

We expand on this body of work, by analyzing the dynamics of a Brownian particle in a tilted cosine potential and computing the intermediate scattering function (ISF). From the ISF we derive time-dependent correlation measures, including the mean-squared displacement, the time-dependent diffusion coefficient, the skewness, and the non-Gaussian parameter. Our work extends Ref.~\cite{Rusch_2025}, which solved the system of a simple, untilted cosine potential, by including an external tilt and quantifying its influence throughout.

The ISF captures how density fluctuations evolve and provides a direct measure of the spatio-temporal correlations of particle motion. Further, it is a central observable in experimental techniques such as differential dynamic microscopy, dynamic light scattering, neutron scattering, and X-ray photon correlation spectroscopy~\cite{ cerbino2008differential,berne2000dynamic,Jeffries_2021,Sinha_2014}. 

The periodicity of the potential allows incorporating an additional wave vector in the standard ISF, leading to a generalized formulation of the ISF. To achieve this, we employ Bloch's theorem, which is a standard tool for analyzing periodic systems, and a spectral-theoretical approach to solve the Fokker-Planck equation.  We obtain analytic expressions of the ISF and the probability density function in terms of eigenfunctions and eigenvalues of the system, which can be computed numerically.

In addition to the ISF, we explore several key dynamical quantities that characterize the particle's motion. We derive the low-order moments using time-dependent perturbation theory. The mean-square displacement and the time-dependent diffusivity provide insights into the influence of the periodic potential and the applied force on the particle's dynamics. We also analyze deviations from normal diffusion by computing the skewness, and the non-Gaussian parameter. Throughout the whole analysis, we focus on analyzing different strengths of the tilting force. In particular, we reproduce the mean velocity~\cite{CHENG_2015} and the effective diffusion to show giant diffusion~\cite{Iida_2025,Reimann_2001,Reimann_2002}.
To validate our analytical results, we perform Brownian dynamics simulations and compare the outcomes with theoretical predictions. Furthermore, we employ a harmonic approximation and solve the deterministic limit in the absence of noise to interpret the system's behavior for small or large tilts, respectively.

The work is organized as follows. In \cref{sec:theory}, we introduce the model system, and the corresponding equations of motion, including the Langevin and Fokker-Planck equations. We solve the Fokker-Planck equation and derive the corresponding observables of interest, the ISF and low-order moments. In \cref{sec:results}, we present and analyze our results, compare simulations with the theoretical predictions, and discuss their physical implications. We conclude in \cref{sec:conclusion} with a summary of the main findings and an outlook on directions for future research.

\section{Theory and Observables} \label{sec:theory}
In this section, we introduce the model and the corresponding equations of motion in both the Langevin and Fokker-Planck formalisms. We solve the Fokker-Planck equation and reformulate it in Dirac notation using the Bloch representation of its eigenfunctions. The work builds on the theoretical framework developed in Refs.~\cite{Chepizhko_2022,Rusch_2024,Rusch_2025}.
The main extension compared to Ref.~\cite{Rusch_2025} is the inclusion of a constant tilt, which makes the operators non-Hermitian and requires distinguishing left and right eigenstates, similar to Refs.~\cite{Chepizhko_2022,Rusch_2024}. 
For completeness, we present all necessary definitions and concepts in a self-contained manner and generalize them to incorporate the periodic structure, going beyond Refs.~\cite{Chepizhko_2022,Rusch_2024}.

Then, we introduce the generalized ISF and low-order moments, such as the variance, time-dependent diffusivity, skewness, and non-Gaussian parameter, adapting the derivation of Ref.~\cite{Rusch_2025} to this setting, by including a non-zero mean drift. We reproduce to the established results, namely the band structure~\cite{Alamilla_2020}, mean drift velocity~\cite{CHENG_2015}, mobility~\cite{Costantini_1999, Sakaguchi_2006, Blickle_2007, Hayashi_2004} and long-time diffusion coefficient~\cite{Iida_2025,Reimann_2001,Reimann_2002} using our method.

\subsection{Model}\label{sec:Model}
The Brownian particle is immersed in a one-dimensional periodic potential $U_0(x) = U_0(x+L) $ tilted by the force $F$. 
We specialize to the tilted cosine form
\begin{align} \label{eq:cosine}
    U(x)= U_1 \cos (Q_1 x)- F x,
\end{align} 
where $Q_1=2\pi/L$ is the wave vector associated with the spatial period $L$.  For convenience, we introduce the dimensionless amplitude $u=U_1/k_BT>0$, and dimensionless force $f=F/k_BT Q_1$ where $k_B T$ is the thermal energy scale. By spatial inversion symmetry one may restrict to $f>0$ without loss of generality.

The equation of motion for the Brownian particle in the presence of the spatially periodic force $-\partial_x U(x) $ is provided by the Langevin equation 
\begin{align} \label{eq:Langevin}
	\dot{x}(t)
	&=  D Q_1 [ u \sin(Q_1 x(t))+f] + \eta(t),	
\end{align}
and yields the time-dependent trajectories $x(t)$.  
The parameter $D$ denotes the short-time diffusion coefficient, with the particle mobility $D/k_B T$ as defined via the Einstein relation. The noise term $\eta (t)$ is a centered Gaussian white noise with delta-correlated variance given by $\langle \eta (t) \eta (t^\prime) \rangle = 2 D \delta (t-t^\prime)$ and zero mean.
We identify three key quantities $D, L$ and $k_BT$ that define the characteristic units of the system. The fundamental length scale is set by the period $L$. The characteristic time scale is determined by the diffusion coefficient $D$, corresponding to the time a free particle takes to diffuse over a period, which is  $L^2/D$. The thermal energy $k_B T$ defines the natural energy scale, leaving two dimensionless control parameters, the dimensionless amplitude $u$ and dimensionless force $f$.

\subsection{Fokker-Planck  equation}\label{sec:Fokker-Planck}

An equivalent, ensemble level description of the Langevin dynamics is provided by the Fokker-Planck equation for the time evolution of the propagator $ \mathbb{P}(x,t| x_0) $.   The propagator is the conditional probability to find the particle at position $ x $ at time $ t $ given it started at $ x_0$ at time zero, with initial condition $\mathbb{P}(x,0| x_0)=\delta(x-x_0)$. We reserve the term Smoluchowski equation for overdamped equilibrium dynamics and refer to the present evolution equation as the Fokker-Planck equation. Using standard methods~\cite{risken1996fokker}, one obtains
\begin{align} \label{eq:FP}
	\partial_t \mathbb{P} 
    	= - D  Q_1 \partial_x [   (u\sin(Q_1 x)+  f ) \mathbb{P} ] + D \partial_x^2 \mathbb{P} =: \mathcal{L}_{0} \mathbb{P}.
\end{align}

To facilitate analytical progress, we replace the infinite line by a finite ring composed of $N\in\mathbb{N}$ unit cells of length $L$, i.e., a domain of size $NL$ with periodic boundary conditions. All results are obtained on this compact domain and the limit $N\to\infty$ is taken at the end. On the ring, nontrivial solutions of \cref{eq:FP} admit separation of variables,
\begin{align}
\mathbb{P}(x,t| x_0) = E(t) \psi(x),
\end{align}
with a purely exponential time dependence $E(t)=e^{-\lambda t}$ and a position-dependent factor $\psi(x)$ that solves an eigenvalue problem. Because the Fokker-Planck operator $\mathcal{L}_0$ is non-Hermitian, we introduce right and left eigenfunctions,
\begin{align}\label{eq:eigenvalue_FP_R}
\mathcal{L}_{0}\psi^{R}_{\lambda}(x) = -\lambda \psi^{R}_{\lambda}(x), \quad
\mathcal{L}_{0}^{\dagger}\psi^{L}_{\lambda}(x) = -\lambda^{*} \psi^{L}_{\lambda}(x),
\end{align}
where $\mathcal{L}_{0}^{\dagger}$ is the adjoint with respect to the scalar product
\begin{align}\label{eq:scalar_product}
\langle \phi | \psi \rangle \coloneq \frac{1}{N}\int_{0}^{NL} \phi(x)^{*} \psi(x) \diff x.
\end{align}
Left and right eigenfunctions can be chosen orthonormal, 
\begin{align} \label{eq:orthonormality_FP}
	\langle \psi^L_\lambda |\psi^R_{\lambda'} \rangle = \frac{1}{N} \int_0^{NL} \psi_\lambda^L(x)^* \psi_{\lambda'}^R(x)   \diff x = \delta_{\lambda\lambda'} ,
	\end{align}
and the eigenfunctions are complete, fulfilling the condition
	\begin{align} \label{eq:completeness_FP}
		\frac{1}{N} \sum_{\lambda} \psi_\lambda^R(x)   \psi_\lambda^L(x_0)^*   = \delta(x-x_0) .
	\end{align} 
For the stationary solution of \cref{eq:FP}, the left-hand side is zero and therefore yielding \cref{eq:eigenvalue_FP_R} for the eigenvalue zero. We can conclude therefore that the right eigenfunction $\psi_{0}^R(x) \propto p^\text{st}(x) $ is proportional to the stationary solution~\cite{Xiao_2015,stratonovich_1967_topics}
\begin{align}\label{eq:stationary_solution}
p^{\mathrm{st}}(x)
= \frac{I_-(x)}{\int_0^L I_-(y)  \mathrm{d}y},
\end{align}
which is normalized for a single cell and the auxiliary function is defined as
\begin{align} \label{eq:aux}
I_{-}(x)
\coloneq \int_0^L e^{- \beta [U(x) - U(x + y)]}   \mathrm{d}y.
\end{align}
We immediately find the left eigenfunction up to a normalization factor by inserting the finding into  \cref{eq:orthonormality_FP} and use the fact that the stationary solution has to be normalized. We choose the normalization factor of the eigenfunctions to be 
\begin{align} \label{eq:stationary_SE}
	\psi_0^R(x) = p^\text{st}(x), \qquad  \psi_0^L(x)=1.
\end{align} 
The formal solution of the Fokker-Planck equation, \cref{eq:FP}, takes the form $\mathbb{P}=e^{\mathcal L_0 t}\delta(x-x_0)$ and by inserting the completeness relation, \cref{eq:completeness_FP}, and applying the eigenvalue equation, \cref{eq:eigenvalue_FP_R}, we obtain
\begin{align} \label{eq:probability_density_FP}
\mathbb{P}(x, t| x_0)  = \frac{1}{N}  \sum_{\lambda} e^{- \lambda t} \psi_\lambda^R(x)  \psi_\lambda^L(x_0)^*.
\end{align}

\subsection{Bloch representation}
Exploiting the spatial periodicity of the force, we decompose eigenfunctions into Bloch waves characterized by a discrete wave vector $q \in (2\pi /NL) \mathbb{Z}$ within the first Brillouin zone $\text{BZ} \coloneq \{  -\pi/L < q \leq \pi/L\}  $, and a band index $n$. With periodic boundary conditions on the ring of length $NL$, the right and left eigenfunctions take the Bloch form
\begin{align}\label{eq:Bloch_ansatz}
\psi^R_{nq}(x) &= e^{i q x} u^R_{nq}(x), \nonumber \\
\psi^L_{nq}(x) &= e^{i q x}u^L_{nq}(x),
\end{align}
where the Bloch amplitudes are periodic, $u^{L/R}_{nq}(x)=u^{L/R}_{nq}(x+L)$.

For each fixed $q$, the biorthonormality of Bloch functions on a unit cell reads~\cite{Rusch_2025}
\begin{align}\label{eq:ortho}
\int_{0}^{L}\!\diff x\,u^L_{nq}(x)^{*} u^R_{mq}(x) = \delta_{nm},
\end{align}
and the completeness relation within a unit cell is
\begin{align}\label{eq:completeness_u}
\delta(x-x_0) = \sum_{n} u^L_{nq}(x_0)^{*} u^R_{nq}(x),
\end{align}
for positions $x$ and $x_0$ in the same cell. 
\begin{figure}[ht!]
    \centering
    \includegraphics[width=0.9\linewidth]{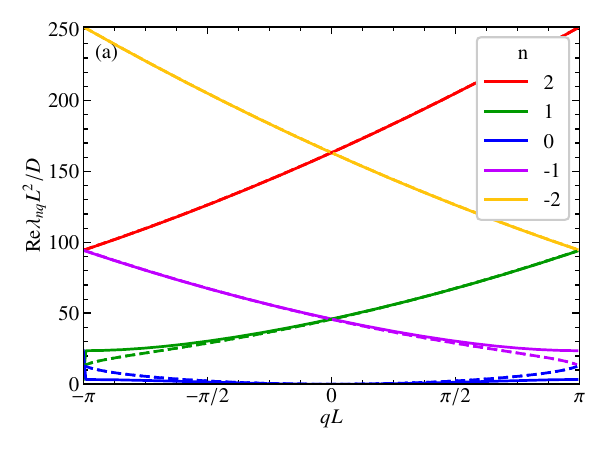}
    \includegraphics[width=0.9\linewidth]{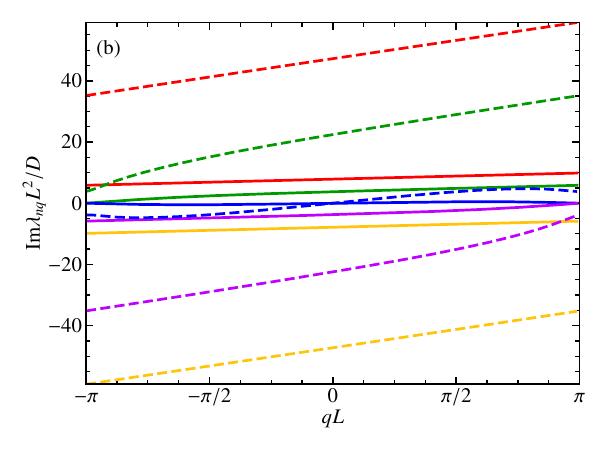}
    \caption{ Real (a) and  imaginary (b) parts of the band structure $\lambda_{nq}$ of the operator $\mathcal L_q$ for the reduced amplitude $u=1$ as a function of the wave vector $q$ and index $n$. Full lines correspond to $f/u=0.1$ and dashed lines to $f/u=0.6$. }
    \label{fig:bands}
\end{figure}
Substituting the Bloch forms into the spectral representation of the propagator, \cref{eq:probability_density_FP}, we obtain
\begin{align}\label{eq:probability_density_S}
\mathbb{P}(x,t | x_0)
= \frac{1}{N}\sum_{q\in\mathrm{BZ}}\sum_{n} e^{-\lambda_{nq} t} e^{i q (x-x_0)} u^R_{nq}(x) u^L_{nq}(x_0)^{*}.
\end{align}

In the large-system-size limit $N\to\infty$, the sum over $q$ becomes an integral over the Brillouin zone,
 \begin{align}
\frac{1}{N}\sum_{q\in\mathrm{BZ}} (\cdots)\ \to\ \frac{L}{2\pi}\int_{\mathrm{BZ}} (\cdots) \mathrm{d}q,
\end{align}
and the eigenvalues $\lambda_{nq}$ form bands labeled by $n\in\mathbb{Z}$, see \cref{fig:bands}. For non-zero tilt the spectrum comes in complex-conjugate pairs, see \cref{eq:spectrum}. We label the bands such that $\mathrm{Im}\,\lambda_{nq}$ is continuous in $q$ and complex conjugation acts as $(n,q)\mapsto(-n,-q)$, implying
 \begin{align} \label{eq:complex_pairs}
u^{R/L}_{nq}(x)=u^{R/L}_{-n,-q}(x)^*,\qquad
\lambda_{nq}=\lambda_{-n,-q}^*.
\end{align}
Comparison with \cref{eq:probability_density_S} shows that the terms occur in complex‑conjugate pairs, thereby ensuring a real-valued probability density.
The eigenvalues are  comprehensively analyzed in Ref.~\cite{Alamilla_2020}. Qualitatively, we find $\lambda_{00}=0$, $\mathrm{Re}\,\lambda_{nq}\ge 0$ for all $n,q$ and higher bands are shifted to more positive values, reflecting faster relaxation rates. As the tilt $f$ increases, the gaps between bands decrease and tend to almost close at the Brillouin-zone edge $qL=\pm\pi$.
At $q=0$ the real parts cross whereas the imaginary parts are opposite in sign. In the vicinity of $q=0$, the slope of the imaginary part of the lowest band  gives the drift velocity~\cite{Challis_2019},
 \begin{align}\label{eq:lowerband_v}
v=\left.\partial_q \mathrm{Im} \,\lambda_{0q}\right|_{q=0},
\end{align}
while the real part controls the diffusion coefficient~\cite{Challis_2019}
 \begin{align} \label{eq:lowerband_D}
D_{\infty}=\left.\tfrac{1}{2} \partial_q^2 \mathrm{Re}\, \lambda_{0q}\right|_{q=0}.
\end{align}
Inserting the above relations into the long-wavelength expansion of the lowest band yields
\begin{align}
\lambda_{0q}
&= \lambda_{00}
  + \left.\partial_q \lambda_{0q}\right|_{q=0} q
  + \frac{1}{2}\left.\partial_q^2 \lambda_{0q}\right|_{q=0} q^2
  + O(q^3) \nonumber \\
  &=  i v q+D_{\infty}q^2  + O(q^3),
\end{align}
where in the second line we used that the real part of $\lambda_{0q}$ is even in $q$ and the imaginary part is odd in $q$, which follows directly from \cref{eq:complex_pairs}.

\subsection{Dirac Notation}
The abstract Dirac notation provides a compact representation and uses the isomorphism between periodic square integrable functions $u^R_{nq}(x), u^L_{nq}(x)\in L^2([0,L])$ and abstract kets $|r_{nq}\rangle$ and bras $\langle l_{nq}|$ in a separable Hilbert space $\mathcal{H}$. Introducing the generalized position basis $|x\rangle$, the Bloch functions are written as $u^R_{nq}(x)=\langle x|r_{nq}\rangle$ and $u^L_{nq}(x)^{*}=\langle l_{nq}|x\rangle$. For fixed $q$, these eigenstates are biorthonormal 
\begin{align}\label{eq:k_orthogonality}
\langle l_{nq}|r_{mq}\rangle = \int_{0}^{L}\diff x\,\langle l_{nq}|x\rangle\langle x|r_{mq}\rangle = \delta_{nm},
\end{align}
and complete
\begin{align}\label{eq:completness_u}
\sum_{n} |r_{nq}\rangle\langle l_{nq}| = \mathbbm{1}.
\end{align}
The position kets are overcomplete on $[0,L]$
\begin{align}
\int_{0}^{L}\diff x\,|x\rangle\langle x| = \mathbbm{1},
\end{align}
and orthogonal in sense of  a Dirac delta function
\begin{align}\label{eq:delta1}
\delta(x-x_0) = \sum_{n}\sum_{q\in\mathrm{BZ}} \langle x|r_{nq}\rangle \langle l_{nq}|x_0\rangle = \langle x|x_0\rangle.
\end{align}
For later calculations it is convenient to use a Fourier basis $\{|\nu\rangle:\nu\in\mathbb{Z}\}$ with real-space representation 
\begin{align} \label{eq:real_space_rep}
  \langle x|\nu\rangle = e^{i Q_{\nu} x}/\sqrt{L},
\end{align}
where $Q_{\nu}\coloneq 2\pi\nu/L$. The Bloch functions then admit the Fourier expansion
\begin{align}\label{eq:Fourierexpansion_p}
u^R_{nq}(x) = \langle x|r_{nq}\rangle = \sum_{\nu\in\mathbb{Z}} \langle x|\nu\rangle \langle \nu|r_{nq}\rangle,
\end{align}
with Fourier coefficients
\begin{align}\label{eq:fourier_coefficients}
\langle \nu|r_{nq}\rangle = \int_{0}^{L}\diff x\, \frac{e^{-i Q_{\nu} x}}{\sqrt{L}} u^R_{nq}(x).
\end{align}

\subsection{Intermediate scattering function}\label{sec:Intermediate_scattering_function}
It is convenient to analyze the dynamics in Fourier space, where the self ISF provides full spatio-temporal resolution and quantifies single particle time correlations for a given wave vector.

With the Bravais lattice $\Lambda=\{nL:\,n\in\mathbb{Z}\}$ and reciprocal lattice $\Lambda^*=\{Q_\mu=2\pi\mu/L:\,\mu\in\mathbb{Z}\}$, any wave vector decomposes uniquely as $k=q+Q_\mu$ with $q\in\mathrm{BZ}, Q_\mu \in \Lambda^*$. We define the generalized ISF as the time correlation of two single particle density modes differing by a reciprocal lattice vector,
\begin{align}\label{eq:ISF_def}
F_{\mu\nu}(q,t)\coloneq \langle e^{-i(q+Q_\mu)x(t)} e^{i(q+Q_\nu)x(0)} \rangle.
\end{align}
For $\mu=\nu$ this reduces to the conventional ISF. Writing the ensemble average as an integral, we obtain
\begin{align}\label{eq:ISF_integral}
F_{\mu\nu}(q,t) =& \int_{0}^{NL}\! \diff x \int_{0}^{L} \! \diff x_0\, e^{-i(q+Q_\mu)x} e^{i(q+Q_\nu)x_0}\nonumber \\
&\times\mathbb{P}(x,t | x_0) p^{\text{st}}(x_0),
\end{align}
assuming the initial position is drawn from the stationary distribution within a single cell. The real-space propagator,  \cref{eq:probability_density_S}, can be obtained by an inverse transform of the generalized ISF~\cite{Rusch_2025}. Inserting \cref{eq:probability_density_S,eq:stationary_SE} and rearranging terms gives~\cite{Rusch_2025}
\begin{align}\label{eq:ISF_integral_short}
F_{\mu\nu}(q,t) =& \sum_{n} e^{-\lambda_{nq} t} \left[\int_{0}^{L}\diff x\, e^{-i Q_\mu x}  u^L_{00}(x)^* u^R_{nq}(x)\right] \nonumber \\
&\times
\left[\int_{0}^{L}\diff x_0\, e^{i Q_\nu x_0}  u^L_{nq}(x_0)^* u^R_{00}(x_0) \right].
\end{align}
To compute $u^{R}_{nq}(x)$ and $u^{L}_{nq}(x)$, we insert the Bloch ansatz \cref{eq:Bloch_ansatz} into \cref{eq:eigenvalue_FP_R}, yielding
\begin{align}
\mathcal{L}_q u^R_{nq}(x) &= -\lambda_{nq} u^R_{nq}(x), \nonumber \\
\mathcal{L}_q^\dagger u^L_{nq}(x) &= -\lambda_{nq}^* u^L_{nq}(x),
\end{align}
with $\mathcal{L}_q=\mathcal{L}_0+\delta\mathcal{L}_q$, where in real space representation
\begin{align}\label{eq:operatorLk}
\mathcal{L}_0 &\coloneq -D Q_1 \partial_x\big[u\sin(Q_1 x)+f\big] + D \partial_x^2, \nonumber\\
\delta\mathcal{L}_q &\coloneq -D Q_1 i q \big[u\sin(Q_1 x)+f\big] + 2 i q D \partial_x - D q^2.
\end{align}
Here the partial derivative acts on everything to its right.
In the Fourier basis, the matrix elements read
 \begin{align}\label{eq:matrix_elements} 
&\langle \mu | \mathcal{L}_0 | \nu \rangle  =  \int_0^{L} \! \frac{\diff x}{L} e^{-i Q_\mu x} \mathcal{L}_0 e^{i Q_\nu x}   \\
&=   -   D  Q_1^2 \left[ \frac{u\mu}{2}(\delta_{\mu, \nu+1}-\delta_{\mu, \nu-1}) + \delta_{\mu, \nu}\left(i \mu f +\mu^2\right) \right]\nonumber,
\end{align}
and
 \begin{align}\label{eq:deltaL_k_matrix} 
\langle \mu | \delta \mathcal{L}_q  | \nu \rangle  =&  
 -DQ_1 q \left[  \frac{u}{2}(\delta_{\mu, \nu+1}-\delta_{\mu, \nu-1})+\delta_{\mu, \nu}( if  +2   \mu )\right] \nonumber \\ 
&-D q^2 \delta_{\mu, \nu} .
\end{align}
This also yields the eigenvalue equation in compact Dirac notation for the right eigenvector $\mathcal{L}_q |r_{nq} \rangle = -\lambda_{nq} |r_{nq} \rangle$, and equivalently for the left eigenvector  $\mathcal{L}_q^\dagger |l_{nq} \rangle = -\lambda_{nq}^* |l_{nq} \rangle $. We numerically find the right and left eigenvectors then by diagonalizing the (truncated) matrix
\begin{align} \label{eq:operator_deltaLk}
	\sum_{\nu \in \mathbb{Z}} \langle \mu | \mathcal{L}_q | \nu \rangle \langle \nu | r_{nq}  \rangle &= -\lambda_{nq} \langle \mu |r_{nq}  \rangle. \end{align}
The spectral representation of the ISF in Dirac notation becomes then
\begin{align} \label{eq:ISF_braket}
	F_{\mu \nu}(q,t) =&	\sum_{n}  e^{-\lambda_{nq}t}  	\\
    & \times \sum_{\sigma, \tau \in \mathbb{Z}} \langle l_{00}|\sigma \rangle  \langle \sigma+\mu | r_{nq}  \rangle \langle l_{nq}|\tau+\nu  \rangle    \langle \tau | r_{00}  \rangle  , \nonumber
\end{align}   
while the conventional ISF with wave vector in the BZ $F(q,t)\coloneqq~F_{00}(q,t)$, simplifies to 
\begin{align} \label{eq:ISF_braket_0}
	F(q,t) 
		&=  \sum_{n}  e^{-\lambda_{nq}t} \langle l_{00}| r_{nq}  \rangle \langle l_{nq}| r_{00}  \rangle =  \langle l_{00}| e^{\mathcal{L}_q t} | r_{00}  \rangle .
\end{align}
 The  stationary state corresponds to  $\lambda_{00}=0$. We also note that the complex-conjugate pairing of the spectrum leads to oscillations in the ISF.

The short-time limit follows from \cref{eq:ISF_def} at $t\to 0$ and \cref{eq:fourier_coefficients}
\begin{align}\label{eq:init_ISF}
&F_{\mu\nu}(q,0) = \big\langle e^{- i Q_{\mu-\nu} x(0)} \big\rangle 
= \int_0^L e^{- i Q_{\mu-\nu} x} p^{\text{st}}(x)\mathrm{d}x\nonumber \\
&= \sqrt{L}\langle \mu-\nu | r_{00} \rangle.
\end{align}
In the long-time limit at $q=0$, temporal correlations do not decay to zero due to the periodic modulation and ISF factorizes
\begin{align}\label{eq:long_time_limit}
F_{\mu\nu}(0,t \to \infty)
&= \big\langle e^{-i Q_\mu x(t)} \big\rangle \big\langle e^{i Q_\nu x(0)}\big\rangle
\nonumber\\
&= L\langle \mu | r_{00} \rangle \langle r_{00}|\nu \rangle.
\end{align}
Equivalently, starting from \cref{eq:ISF_braket} one recovers the same result by inserting $\langle l_{00} | \sigma\rangle = \sqrt{L}\delta_{\sigma,0}$.

The Fourier coefficients entering the above expressions can be obtained directly in real space, without solving the full eigenvalue problem
\begin{align}
\langle \mu | r_{00} \rangle
= \frac{1}{\sqrt{L}}
\frac{\sum\nolimits_{\nu=-\infty}^\infty I_{\mu-\nu}(-u) I_\nu(u)B_\nu}
{\sum\nolimits_{\sigma=-\infty}^{\infty} I_\sigma(u)^2B_\sigma},
\end{align}
where $I_n(\cdot)$ is the modified Bessel function of the first kind of integer order $n$  and
\begin{align} 
B_\nu = \frac{\exp\big[(i \nu - f) Q_1 L\big]-1}{(i \nu - f)Q_1}.
\end{align}
The derivation is provided in \cref{sec:real_space_coefficients}. In the limit of vanishing tilt $f\to 0$, this reduces to the known result~\cite{Rusch_2025}
\begin{align}
\langle \mu | r_{00} \rangle = \frac{(-1)^\mu I_\mu(u)}{\sqrt{L} I_0(u)}.
\end{align}

By definition, the ISF obeys  space-inversion symmetry
\begin{align}\label{eq:space_inversion}
F_{\mu\nu}(q,t) = F_{-\mu, -\nu}(-q,t)^*,
\end{align}
and at the Brillouin-zone edge it satisfies the additional relation~\cite{Rusch_2025}
\begin{align}\label{eq:BZ_edge}
F_{\mu\nu}\left(\frac{\pi}{L},t\right)
= F_{\mu+1, \nu+1}\left(-\frac{\pi}{L}, t\right)
= F_{-(\mu+1),-(\nu+1)}\left(\frac{\pi}{L},t\right)^*.
\end{align}

\subsection{Low-order cumulants} \label{sec:low_order_moments}
The diagonal ISF, $F(q,t)$, is the characteristic function of the random displacements $\Delta x(t)$. As there is a tilt in our system, we are interested in the cumulants, which can be derived by the cumulant expansion of the ISF for small wave vectors
\begin{align} \label{eq:cumulant_expansion}
    \ln F(q, t)=\sum_{j=1}^{\infty} \frac{(-\mathrm{i} q)^j \kappa_j[\Delta x(t)]}{j!},
\end{align}
with $\kappa_j[\Delta x(t)]$ being the $j$th cumulant.
It is convenient to decompose the operator $\delta \mathcal L_q $, \cref{eq:operatorLk}, as 
\begin{align}
    \delta \mathcal L_q = \delta \mathcal L_q^1-Dq^2,
\end{align} 
where $\delta \mathcal L_q^1$ denotes the part linear in $q$. Accordingly, the ISF can be written as
\begin{align} \label{eq:qseparation_ISF}
    F(q,t) &= e^{-D q^2 t} \tilde{F}(q,t), 
\end{align} 
which introduces the reduced ISF with reduced time-evolution operator linear in $q$,
\begin{align} \label{eq:reduced_ISF}
 \tilde{F}(q,t) &\coloneq \langle l_{00}|e^{(\mathcal{L}_0+\delta \mathcal{L}_{q}^1 ) t }  | r_{00}  \rangle.
\end{align} 

Finally, by expanding the reduced ISF, \cref{eq:reduced_ISF}, in $q$ and comparing it term-by-term with \cref{eq:cumulant_expansion}, the cumulants can be read off directly. The expansion of the reduced ISF is presented in the following, after a few remarks.
Employing the \emph{reduced} ISF simplifies this procedure substantially. The linear dependence on $q$  of the reduced time-evolution operator  precludes mixed terms in the expansion, so each cumulant is determined directly by its coefficient. The full cumulants are trivially obtained from the reduced $\tilde\kappa_j[\Delta x(t)]$ as they coincide with those of the conventional ISF, except for the second one 
\begin{align} \label{eq:tranform_red}
\kappa_j\!\left[\Delta x(t)\right]
=
\tilde{\kappa}_j\!\left[\Delta x(t)\right]
+
2Dt \delta_{j,2}.
\end{align}

The derivation of the ISF expansion is described in detail in Refs.~\cite{Rusch_2024,Chepizhko_2022}. The only difference here is that we use the reduced expansion and the reduced ISF, following~\cite{Rusch_2025}. For completeness, we summarize the essential steps. Starting from the explicit expression for the ISF, \cref{eq:reduced_ISF}, we expand it by iteratively substituting the reduced time-evolution operator using the Dyson representation
\begin{align} \label{eq:dyson_representation}
	&e^{(\mathcal{L}_0+\delta \mathcal{L}_{q}^1 ) t } = e^{\mathcal{L}_0 t} + \int_0^t \!\diff s \, e^{\mathcal{L}_0 (t-s)} \delta \mathcal{L}_{q}^1 e^{(\mathcal{L}_0 +\delta \mathcal{L}_q^1 ) s} .
\end{align} 
Inserting this into \cref{eq:reduced_ISF}, the main simplifications follow from $e^{\mathcal{L}_0 t}  | r_{n0} \rangle = | r_{n0} \rangle $ and $  \langle l_{n0} | e^{\mathcal{L}_0 t} = \langle  l_{n0} | $, together with the completeness relation \cref{eq:completness_u} at $q=0$. Taking the logarithm and evaluating the resulting integrals yields the final form
\begin{widetext}
\begin{align} \label{eq:explicit_ISF_cumulant}
\ln \tilde{F}(q, t)= & \; t\left\langle l_{00} | \delta \mathcal{L}_q^{1} r_{00}\right\rangle
+\sum_{n \neq 0} \frac{e^{-\lambda_{n0} t}+\lambda_{n0} t-1}{\lambda_{n0}^2}
\left\langle l_{00} | \delta \mathcal{L}_q^{1} r_{n0}\right\rangle
\left\langle l_{n0} | \delta \mathcal{L}_q^{1} r_{00}\right\rangle \\
& +\sum_{n \neq 0}
\frac{\lambda_{n0} t+e^{-\lambda_{n0} t}\left(\lambda_{n0} t+2\right)-2}{\lambda_{n0}^3}
\left\langle l_{00} | \delta \mathcal{L}_q^{1} r_{n0}\right\rangle
\left\langle l_{n0} | \delta \mathcal{L}_q^{1} r_{00}\right\rangle
\left(
\left\langle l_{n0} | \delta \mathcal{L}_q^{1} r_{n0}\right\rangle
-
\left\langle l_{00} | \delta \mathcal{L}_q^{1} r_{00}\right\rangle
\right) \nonumber \\
& +\sum_{n \neq 0} \sum_{m \neq 0, m \neq n}
\left(
\frac{e^{-\lambda_{n0} t}+\lambda_{n0} t-1}{\lambda_{n0}^2(\lambda_{m0}-\lambda_{n0})}
+\frac{e^{-\lambda_{n0} t}+\lambda_{n0} t-1}{\lambda_{m0}^2(\lambda_{n0}-\lambda_{m0})}
\right)
\left\langle l_{00} | \delta \mathcal{L}_q^{1} r_{n0}\right\rangle
\left\langle l_{n0} | \delta \mathcal{L}_q^{1} r_{m0}\right\rangle
\left\langle l_{m0} | \delta \mathcal{L}_q^{1} r_{00}\right\rangle
+O(q^4), \nonumber
\end{align}

\end{widetext}
with apparent singular terms resolved separately. Accordingly, the sums are restricted to nonzero integers and $m\neq n$, while zero modes and coincident indices are treated by taking limits.

As the expressions for higher moments become lengthy, it is convenient to express the fourth cumulant in terms of moments. All reduced moments up to third order can be obtained from the reduced cumulants, which follow from \cref{eq:explicit_ISF_cumulant}. The derivation of the fourth moment proceeds analogously to the cumulants, except that we expand the ISF in moments,
\begin{align} \label{eq:moment_expansion}
    \tilde{F}(q, t)=\sum_{j=0}^{\infty} \frac{(-\mathrm{i} q)^j \tilde{\mu}_j[\Delta x(t)]}{j!}.
\end{align}
The resulting moment expansion of the reduced ISF is given in the Appendix, \cref{eq:moments}. We then convert the reduced moments back via
\begin{align}
\mu_n[\Delta x(t)] = \sum_{k=0}^{| n/2 |} 
\frac{n!  (D t)^k}{(n-2k)!   k! }    \tilde{\mu}_{ n-2k}[\Delta x(t)]  ,
\end{align}
where $\mu_0[\Delta x(t)] =1$.

The final results for the low-order moments begin with the mean drift velocity, derived from the first cumulant
\begin{align} \label{eq:velocity}
    v= \frac{\text{d}\kappa_1[\Delta x(t)]}{\text{d}t}= \frac{i}{q}\langle l_{00} |   \delta \mathcal{L}_{q}^1   r_{00}\rangle.
\end{align}
The variance is given by
\begin{align}\label{eq:msd}
&\text{Var}[ \Delta x(t) ] = 2Dt \\
&-\frac{2}{q^2} \sum_{n \neq 0} \frac{e^{-\lambda_{n0} t} + \lambda_{n0} t -1}{\lambda_{n0}^2}  \langle l_{00} | \delta \mathcal{L}_{q}^1 | r_{n0} \rangle  \langle l_{n0} | \delta \mathcal{L}_{q}^1 |  r_{00} \rangle  \nonumber ,
\end{align}
and its derivative yields the time-dependent diffusion coefficient  
\begin{align} \label{eq:diffusivity_explicitely}
	&D(t) \coloneqq  \frac{1}{2} \frac{\diff \text{Var}(t)}{\diff t} 
\end{align}
in one dimension. At long times, the system is characterized by the long-time diffusion coefficient $D_\infty~\coloneqq~D(t\to~\infty)$
\begin{align} \label{eq:diffusivity_explicitely_longtime}
	D_\infty =& D-\frac{1}{ q^2 }    \sum_{n \neq 0} \frac{1 }{\lambda_{n0}}  \langle l_{00} | \delta \mathcal{L}_{q}^1 | r_{n0} \rangle 
	 \langle l_{n0} | \delta \mathcal{L}_{q}^1 |  r_{00} \rangle ,
\end{align}
 where an analytic expression for arbitrary (tilted) periodic potentials is known also in real-space representation~\cite{Lifson_1962,Festa_1978,Reimann_2002,Straube_2024,Spiechowicz_2023}.
The skewness and non-Gaussian parameter are computed from the cumulants via
\begin{align} \label{eq:skew_non_gaussian}
    \text{Skew}(t)=\frac{\kappa_3[\Delta x(t)]}{\kappa_2^{2/3}[\Delta x(t)]}, \qquad \alpha_2(t)= \frac{\kappa_4[\Delta x(t)]}{3\kappa_2^2[\Delta x(t)]} .
\end{align}
but are too lengthy to show in full form in the main text. 
The cumulants up to the third order are given in \cref{sec:cumulants} and the fourth order moment for operators linear in $q$ can be found in the Appendix of Ref.~\cite{Rusch_2024}, with minor notational differences.

\subsection{Characteristics of the potential and limiting cases}\label{sec:HA_det}
\begin{figure}[ht!]
    \centering
    \includegraphics[width=0.9\linewidth]{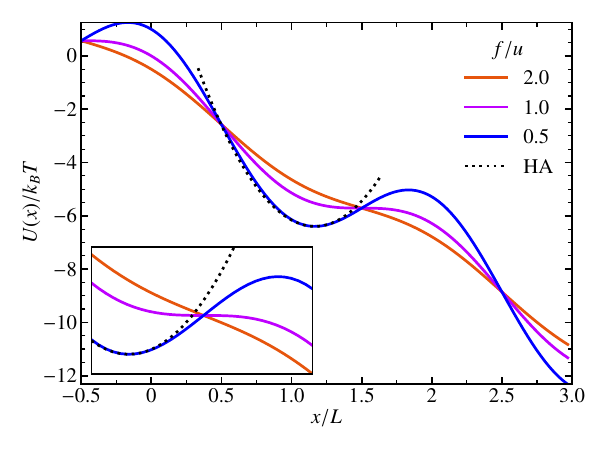}
    \caption{Tilted washboard potential for different ratios of $f/u$: The running state, $f>u$, critical tilt, $f=u$, and locked state, $f<u$. The inset is zoomed to the inflection point, the black dotted line is the harmonic approximation (HA) around the local minima. }
    \label{fig:potential}
\end{figure}

In order to interpret the full results, derived in \cref{sec:Intermediate_scattering_function,sec:low_order_moments}, we analyze the given potential, \cref{eq:cosine}, and solve limiting cases that aid understanding of the dynamics. 
The tilted washboard potential, \cref{eq:cosine}, in dimensionless units
\begin{align} \label{eq:cosine_dimless}
   \beta U(x)=u \cos(Q_1x)-Q_1fx,
\end{align}
with inverse thermal energy $ \beta =1/k_BT$ is governed by the ratio of tilt to amplitude, $f/u$, and the behavior can be distinguished in three regimes. 
In the locked state, $f<u$, local minima and maxima exist and the particle can be (temporarily) trapped in the minima. At the critical tilt, $f=u$, the minimum and its adjacent maximum coalesce into an inflection point, and in the running state, $f>u$, no local extrema remain and motion is unbounded, see \cref{fig:potential}.

For $f<u$ the local minima and local maxima can be found
\begin{align} \label{eq:xmin}
     Q_1 x_{\min}^n&=\pi + \arcsin\!\left(f/u\right) + 2\pi n, \nonumber \\
     Q_1 x_{\max}^n &= -\arcsin\!\left(f/u\right) + 2\pi n, 
\end{align}
with $n\in\mathbb{Z}$. 
Kramers' escape time for the tilted periodic potential is  provided by~\cite{Haenggi_1990,stratonovich_1967_topics,Coffey_2006,MELNIKOV_1991}
\begin{align} \label{eq:tau_esc}
\tau_{\mathrm{esc}} \simeq \frac{1}{k_{+}+k_{-}}= \frac{2\pi e^{\beta \Delta U_{+}}}{Q_1^2D u \sqrt{1-(f/u)^2}} 
\frac{1}{1+e^{-2\pi f}}.
\end{align}
However, this equation requires sufficiently high barriers $u\gg f$.
Near the critical tilt, $f \to u$ for $f<u$, the minimum and the adjacent maximum coalesce in a saddle-node bifurcation and the mean-escape time in this near-threshold regime is~\cite{MELNIKOV_1991,Coffey_2006}
\begin{align} \label{eq:tau_esc_ftou}
\tau_{\mathrm{esc}}^{f \to u}
\;\simeq\; \frac{2\pi}{D Q_1 u \sqrt{1-(f/u)^2}}\;
\exp\!\Big[\frac{2}{3} u \left(1-\left(f/u\right)^2\right)^{3/2}\Big].
\end{align}

\subsubsection{Deterministic limit}
In the running state, where $f>u$, we consider the deterministic limit, which is the noise-free dynamics, $T\to 0$, of the Langevin equation, \cref{eq:Langevin}
\begin{align}
\dot{x}(t) = D Q_1 \bigl[ u \sin(Q_1 x(t)) + f \bigr].
\end{align}
In this regime, the particle is never trapped, as the potential no longer possesses minima, and it drifts continuously. The time required to traverse one period of the potential, $L$, is given by
\begin{align}
T = \int_0^{L} \frac{dx}{\dot{x}} = \frac{1}{D Q_1^2} \int_0^{2\pi} \frac{dy}{f + u \sin y} = \frac{2 \pi}{D Q_1^2 \sqrt{f^2 - u^2}},
\end{align}
where we used the substitution $y = Q_1 x$.  
From this, the mean drift velocity is obtained
\begin{align} \label{eq:vdet}
v_{\text{det}}=\langle \dot{x} \rangle = \frac{L}{T} = D Q_1 \sqrt{f^2 - u^2}.
\end{align}
Since the motion is purely deterministic, all high-order cumulants vanish. 

To compute the mobility we take the derivative with respect to the force $F= k_B T Q_q f$
  \begin{align}
    (\mu_{\infty})_{\text{det}}
    &:= \frac{1}{k_B T Q_1}\,\frac{\partial v_{\text{det}}}{\partial f}= \frac{D}{k_B T} \frac{f/u}{\sqrt{f^2/u^2-1}}.
\end{align}

\subsubsection{Harmonic approximation} \label{app:HA}
In the locked state  $f<u$, the tilted cosine potential can be linearized around its local minimum, \cref{eq:xmin}. If we now consider a small perturbation $\bar x(t)$ around a  minimum, we can write the trajectory as
\begin{align}
    x(t)=x^n_\text{min}+\bar x(t),
\end{align}
and the Langevin equation \cref{eq:Langevin} linearizes to an Ornstein-Uhlenbeck dynamics,
\begin{align}
    \frac{\text{d}}{\text{d}t} \bar x(t)&=-\frac{\bar x(t)}{\tau_\mathrm{HA}}+ \eta(t),
\end{align}
with  harmonic-well relaxation time 
\begin{align} \label{eq:tau_ha}
    \tau_\mathrm{HA} = 1 /D  \sqrt{u^2-f^2} Q_1^2 .
\end{align}

The corresponding Fokker-Planck equation for the fluctuation propagator $\mathbb{P} \coloneq \mathbb{P}(\bar x t |  \bar x_0)$ reads
\begin{align}
\partial_t \mathbb{P} = \partial_{\bar x}  \frac{\bar x \mathbb{P} }{\tau_\mathrm{HA}} + D\partial^2_{\bar x} \mathbb{P}.
\end{align}
The  Ornstein-Uhlenbeck propagator is Gaussian~\cite{risken1996fokker}
\begin{align}
    \mathbb{P}(\bar x t |  \bar x_0)=\frac{1}{\sqrt{2 \pi V(t)}} e^{-\left(\bar x- \bar x_0 e^{-t / \tau_\mathrm{HA}}\right)^2/2 V(t) },
\end{align}
with $V(t)=D \tau_\mathrm{HA}\left[1-\exp{(-2 t / \tau_\mathrm{HA})}\right]$, and its long-time limit yields the stationary distribution
\begin{align}
    p^\text{st}(\bar x_0)=\frac{1}{\sqrt{2 \pi D \tau_\mathrm{HA}}} e^{-\bar x_0^2/2 D \tau_\mathrm{HA}}.
\end{align}
Starting from the ISF, \cref{eq:ISF_integral}, and inserting the propagators, defined above, we find
\begin{align} \label{eq:ISF_integral1}
		F_{\mu\nu}(q,t) =& \int_{-\infty}^{\infty} \! \diff \bar x \int_{-\infty}^{\infty} \! \diff  \bar x_0   e^{-i(q+Q_\mu) (\bar x+x_{\min}^n )}  \nonumber \\
		&\times e^{i(q+Q_\nu) (\bar x_0+x_{\min}^n)}\mathbb{P}(\bar x, t |  \bar x_0 )  p^{\text{st}}( \bar x_0) ,
\end{align}
which evaluates to 
\begin{align}\label{eq:ISF_HA} 
   &F_{\mu \nu}(q,t) \\
   &=  \exp \left[-\frac{D \tau_\mathrm{HA} }{2}  \left[(q+Q_\mu)^2-2 (q+Q_\mu) (q+Q_\nu) e^{-\frac{t}{\tau_\mathrm{HA} }} \right. \right. \nonumber\\
   &\left. \left.+(q+Q_\nu)^2\right]\right](-1)^{\mu-\nu} \; \exp\Big[-i (\mu-\nu) \arcsin(f/u) \Big].\nonumber
\end{align}
From the Ornstein-Uhlenbeck propagator the mean-square displacement is obtained
\begin{align}
    \langle \bar x(t)^2 \rangle &=\int_{-\infty}^{\infty} \! \diff \bar x \int_{-\infty}^{\infty} \! \diff \bar x_0 \,(\bar x-\bar x_0)^2 \times \mathbb{P}(\bar x, t | \bar x_0 )   p^{\text{st}}(\bar x_0) \nonumber \\
    &= 2 D \tau_\mathrm{HA}\left(1-e^{- t / \tau_\mathrm{HA}}\right),
\end{align}
and also the time-dependent diffusivity proportional to the derivative
\begin{align}
    D(t)
    &=  De^{- t / \tau_\mathrm{HA}}.
\end{align}
The mean drift velocity is zero, so are all cumulants except the second one. 

\begin{figure}[tb]
    \centering
    \includegraphics[width=0.9\linewidth]{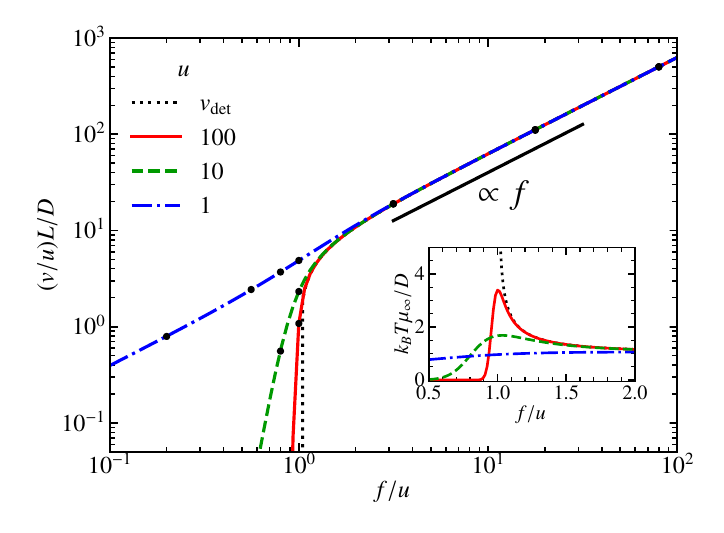}
    \caption{Mean drift velocity $v$ rescaled by $u$ and mobility $\mu_\infty$ (inset) as a function of the tilt-to-amplitude ratio $f/u$ for several potential amplitudes $u$ (colored curves: analytical; black markers: simulations). The black dotted curve shows the deterministic prediction $v_{\mathrm{det}}$. }
    \label{fig:vel_fu}
\end{figure}
\begin{figure}[tb]
    \centering
    \includegraphics[width=0.9\linewidth]{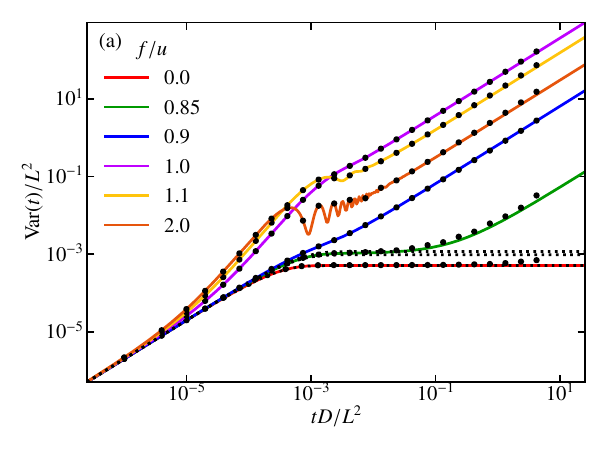}
    \includegraphics[width=0.9\linewidth]{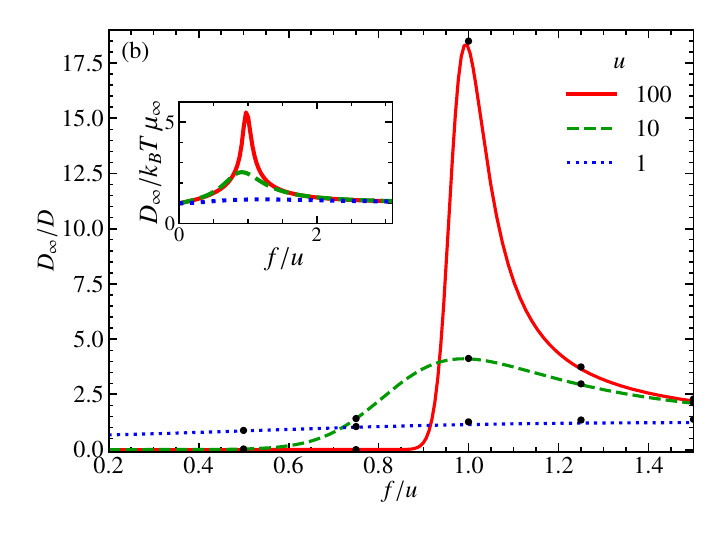}
    \caption{(a) Displacement variance as a function of time for various tilt-to-amplitude ratios $f/u$ at $u=100$. Black dotted curves: harmonic approximation in the locked regime ($f<u$). (b) Long-time diffusion coefficient $D_{\infty}$ versus $f/u$ for different potential amplitudes $u$ (colored curves: analytical; black markers: simulations), highlighting the giant-diffusion peak near $f\simeq u$. Inset: ratio of  mobility and long-time diffusion coefficient as a funcion of $f/u$}
    \label{fig:Var_Dt_fu}
\end{figure}
\begin{figure}[tb]
    \centering
    \includegraphics[width=0.9\linewidth]{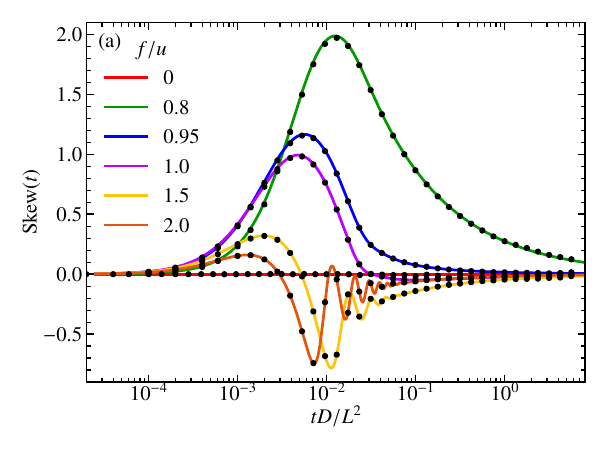}
    \includegraphics[width=0.9\linewidth]{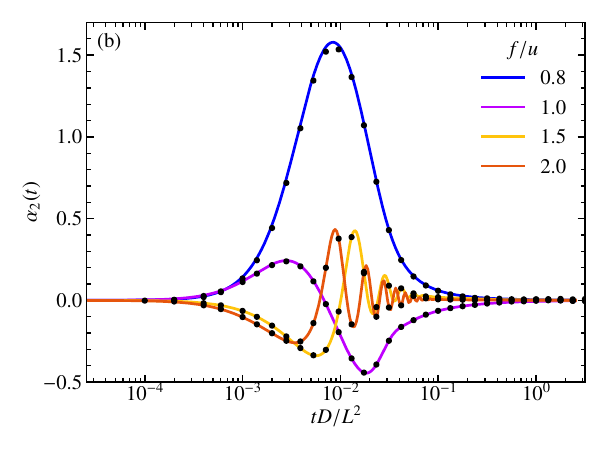}
            \caption{Skewness (a) and non-Gaussian (b) parameter as a function of time $t$ for different tilt ratios $f/u$ and for $u=10$. Colored lines are the analytical results and black markers the simulation results. }
	\label{fig:skewness_nonGaussian}
\end{figure}

\begin{figure*}[htb]
    \centering
    \includegraphics[width=1.0\linewidth]{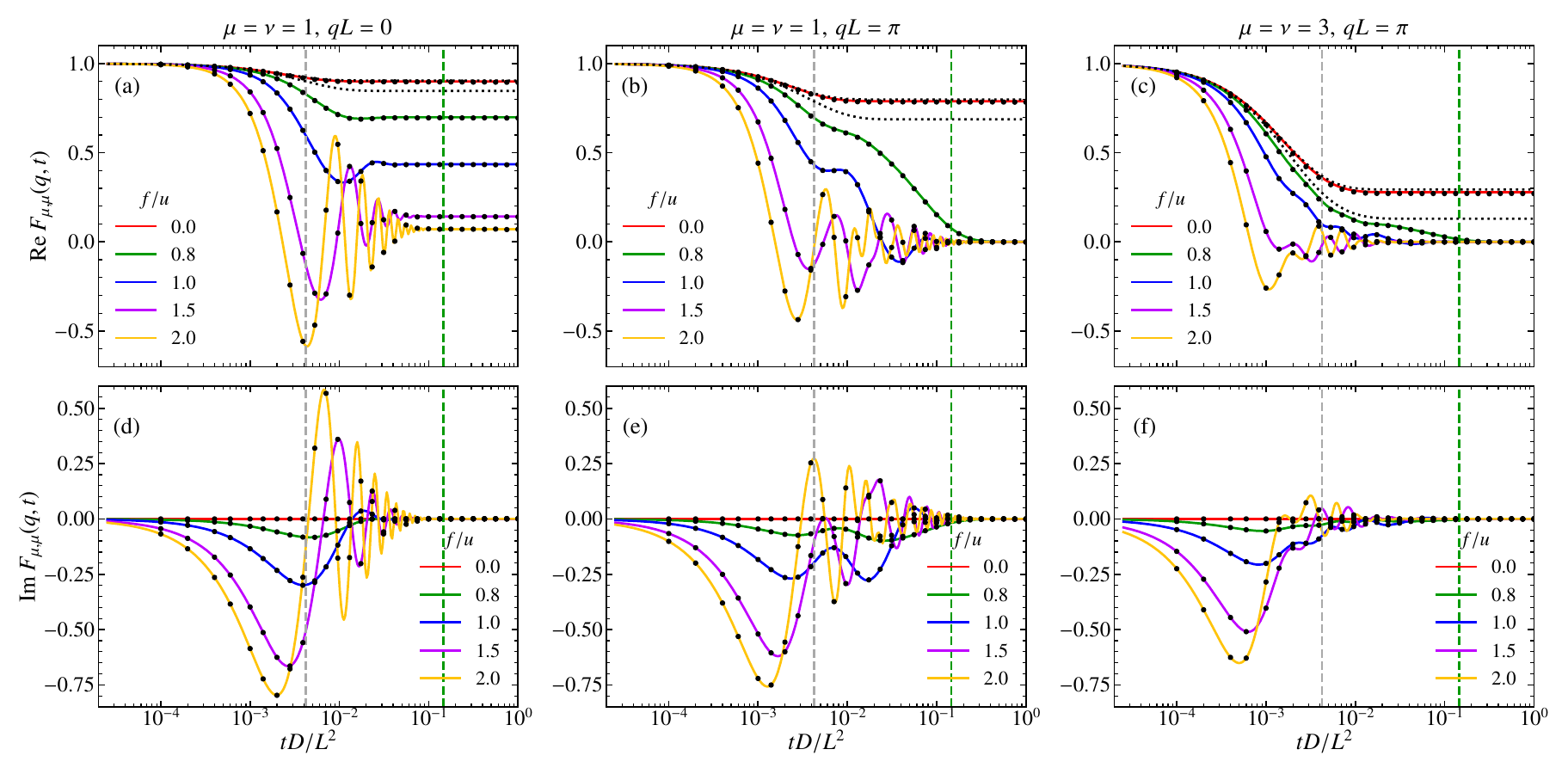}
        \caption{Real and imaginary part of the diagonal ISF, $\mu=\nu$, as a function of time $t$ for different tilt ratios $f/u$ and for $u=10$. Each panel (a-c) and (d-f) shows a different value for the wave vectors $q, Q_\mu$, and $Q_\mu$. The black dotted lines are the values for the HA in the case of $f<u$. The  dashed grey helping line is marking $\tau_\mathrm{HA}$ and the dashed green helping line corresponds to the escape time $\tau_\mathrm{esc}$ for the case of $f/u=2.0$. Colored lines are the analytical results and black markers the simulation results.  }
	\label{fig:ISF_diagonal}
\end{figure*}

\section{Results}\label{sec:results}
In this section, we characterize the overdamped dynamics of a single Brownian particle in a tilted washboard potential by evaluating the observables derived in \cref{sec:Intermediate_scattering_function,sec:low_order_moments}. The mean velocity and long time diffusivity are benchmarked against established results~\cite{CHENG_2015,Reimann_2001,Reimann_2002}. Our analysis goes beyond that of Reimann \emph{et al.}~\cite{Reimann_2001,Reimann_2002} by deriving the full time dependence of the variance and quantifying deviations from Gaussianity via the skewness and the non-Gaussian parameter. We then present diagonal and off-diagonal components of the ISF across tilts and wave vectors, providing a complete spatio-temporal characterization of the dynamics. Throughout, analytical predictions are validated by simulations and interpreted using the characteristic time scales, the deterministic limit, and the harmonic approximation from \cref{sec:HA_det}.

The control parameters are the dimensionless potential amplitude $u=U_1/k_B T$, measuring the modulation strength relative to thermal noise, and the tilt-to-amplitude ratio $f/u$. The critical tilt $f=u$ marks the transition from locked to running dynamics. The most pronounced effects occur in the low-temperature regime $u \gg 1$.

We performed Brownian dynamics simulations of single particles in a one-dimensional tilted cosine potential, integrating the overdamped Langevin equation with the Euler-Maruyama scheme~\cite{maruyama1955continuous}. 
To access long-time behavior on logarithmic time scales, we used the extended order-$n$ algorithm of Frenkel and Smit~\cite{frenkel2002understanding,siems2017computersimulationen,Siems_2018}. In principle, both the simulations and the analytical framework solve the same stochastic dynamics and must yield identical results. Any discrepancies are purely methodological: simulations carry statistical uncertainties from finite sampling and trajectory averaging, whereas the analytical route incurs numerical errors from matrix truncation and the consequent approximations in computed eigenvalues.
\begin{figure*}[htb]
    \centering
    \includegraphics[width=1.0\linewidth]{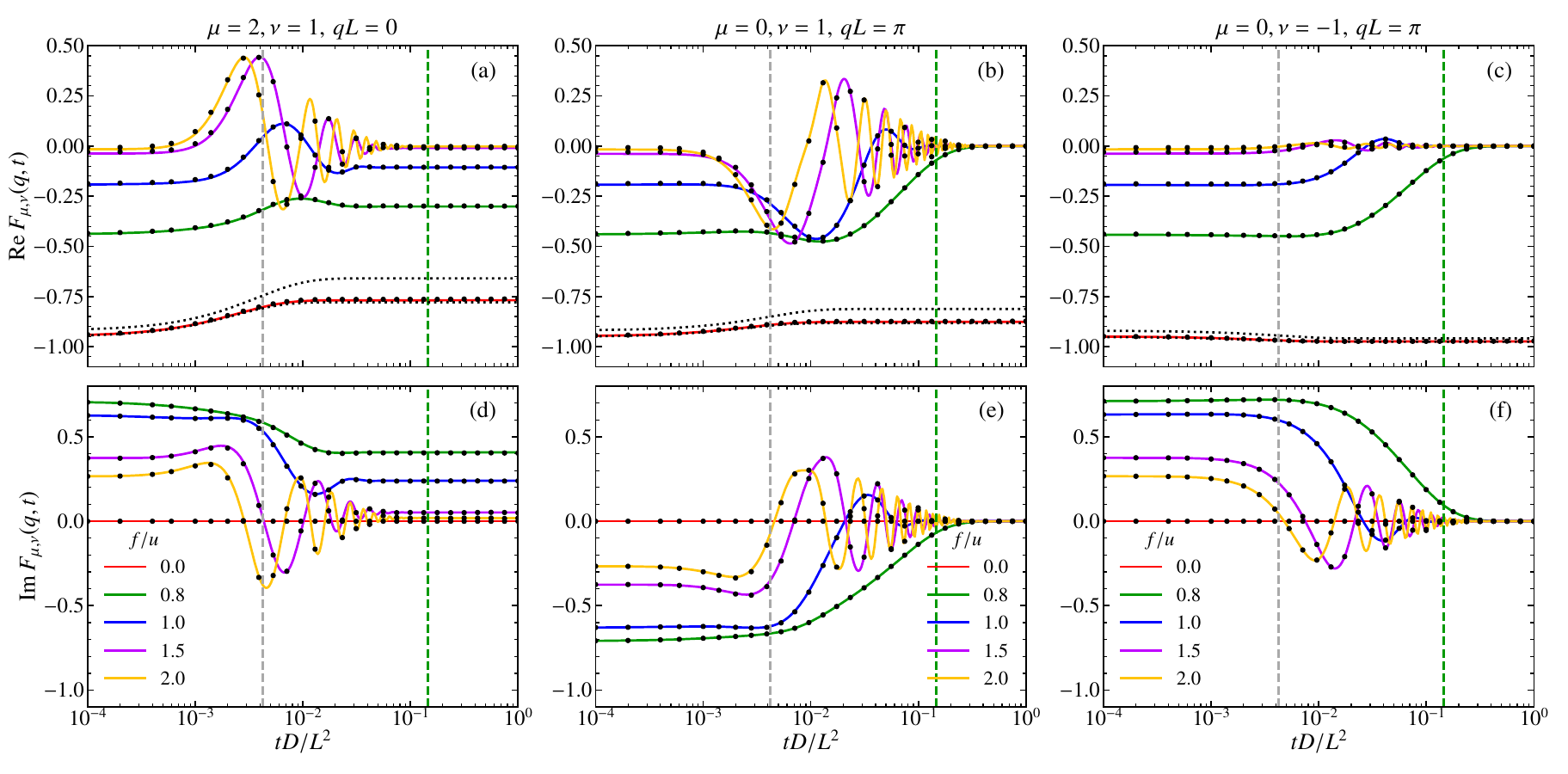}
        \caption{Real and imaginary part of the off-diagonal ISF, $\mu\neq\nu$, as a function of time $t$ for different tilt ratios $f/u$ and for $u=10$. Each panel (a-c) and (d-f) shows a different value for the wave vectors $q, Q_\mu$, and $Q_\mu$. The black dotted lines are the values for the HA in the case of $f<u$. The  dashed grey helping line is marking $\tau_\mathrm{HA}$ and the dashed green helping line corresponds to the escape time $\tau_\mathrm{esc}$ for the case of $f/u=2.0$. Colored lines are the analytical results and black markers the simulation results. }
	\label{fig:ISF_offdiagonal}
\end{figure*}

\subsection{Mean velocity, variance, and diffusivity}
Figure \ref{fig:vel_fu} shows the rescaled mean velocity $v/u$, \cref{eq:velocity}, versus $f/u$ for several amplitudes $u$. In the locked regime the drift is strongly suppressed, and the curves steepen near the critical tilt. For large amplitudes $u\gg1$, the results converge to the deterministic prediction, \cref{eq:vdet}, and the different $u$-curves collapse in the running regime, consistent with Ref.~\cite{CHENG_2015}.

In the locked state, $f<u$, the variance \cref{eq:msd} exhibits two-step growth, see \cref{fig:Var_Dt_fu}~(a): an initial diffusive rise $2Dt$ with intra-well relaxation around $t\simeq\tau_{\mathrm{HA}}$ \cref{eq:tau_ha}, whose plateau height is captured by the harmonic approximation \cref{sec:HA_det}, followed by a long-time diffusive regime with slope $2D_\infty$ that sets in at $t\propto\tau_{\mathrm{esc}}$ \cref{eq:tau_esc}. As $f\to u$, the escape time approaches $\tau_{\mathrm{esc}}^{f\to u}$ \cref{eq:tau_esc_ftou}.

In the running regime, the variance shows oscillations around $t\simeq\tau_{\mathrm{HA}}$ \cref{eq:tau_ha}, originating from the interplay of constant drift and periodic forcing, before entering a linear regime $2D_\infty t$ [\cref{eq:diffusivity_explicitely_longtime}]. At the critical tilt, the long-time growth is maximal, consistent with the giant-diffusion phenomenon, extensively analyzed by Reimann \emph{et al.}~\cite{Reimann_2001,Reimann_2002}.

\Cref{fig:Var_Dt_fu}~(b) reproduces the long-time diffusivity of Refs.~\cite{Reimann_2002,Iida_2025}: the ratio $D_\infty/D$ peaks near $f=u$. With decreasing temperature (increasing $u$), the peak height grows and its width narrows and for high temperatures ($u\ll1$) the enhancement is weak and $D_\infty\approx D$.

In thermal equilibrium, the Einstein relation connects the diffusion coefficient and mobility. 
However, for the system considered here, the long-time transport coefficients do not generally satisfy this relation. Instead, we find
\begin{align} \label{eq:GER}
    D_\infty \neq k_B  T \mu_{\infty} . 
\end{align}
where $\mu_{\infty}$ denotes the mobility and $D_\infty$ the long-time diffusivity given in \cref{eq:diffusivity_explicitely_longtime}. 
We note that at equilibrium, $f=0$ the Einstein relation $D_\infty =k_B  T \mu_{\infty}$ is recovered.
The mobility is obtained by differentiating the velocity, \cref{eq:velocity}, with respect to the tilt $F$ 
\begin{align} \label{eq:mobility_def}
    \mu_{\infty}
    &\coloneq \frac{\partial v}{\partial F}
    = \frac{1}{k_B T Q_1}\,\frac{\partial v}{\partial f},
\end{align}
and reads explicitly
\begin{align} \label{eq:mobility}
    k_B T\mu_{\infty}
    = D
    + \frac{i}{qQ_1}\sum_{n \neq 0}
    \frac{
        \big\langle l_{n0} \big| \tfrac{\partial \mathcal{L}_0}{\partial f} \big| r_{00} \big\rangle\,
        \big\langle l_{00} \big| \delta\mathcal{L}_q^{1} \big| r_{n0} \big\rangle
    }{\lambda_{n0}},
\end{align}
see \cref{sec:GER} for the derivation. The mobility shows features similar to those of the long-time diffusion coefficient, such as an enhancement for high amplitudes $u$ around the bifurcation point, see \cref{fig:vel_fu}. In the literature, real-space representations of  \cref{eq:velocity,eq:diffusivity_explicitely_longtime} as well as discussions of deviations from the Einstein relation,  \cref{eq:GER}, are available~\cite{Costantini_1999, Sakaguchi_2006, Blickle_2007, Hayashi_2004}. For further details, we refer the reader to these references.

\subsection{Skewness and non-Gaussian parameter}
The skewness and non-Gaussian parameter, \cref{eq:skew_non_gaussian}, quantify asymmetry and deviations from Gaussianity in the displacement distribution. For $f<u$, the skewness is positive, peaking below the critical tilt and vanishing at $f=0$ by symmetry, see \cref{fig:skewness_nonGaussian}. This indicates maximal asymmetry when the particle is trapped yet senses the tilt. For $f>u$, the skewness oscillates, reflecting periodic episodes of enhanced forward motion.

The non-Gaussian parameter shows an analogous dependence on $f/u$: it peaks below $f=u$ and becomes oscillatory in the running regime. In both the short-time ($t\lesssim\tau_{\mathrm{HA}}$, \cref{eq:tau_ha}) and long-time limits ($t\gtrsim\tau_{\mathrm{esc}}^{f\to u}$, \cref{eq:tau_esc_ftou}), skewness and the non-Gaussian parameter approach zero, indicating symmetric, Gaussian behavior. Across conditions, simulations agree with the analytical predictions.

\subsection{Intermediate scattering function}\label{sec:results_ISF}

The ISF, \cref{eq:ISF_braket}, captures spatio-temporal correlations of the particle motion and provides the full information on the particle dynamics. We present its real and imaginary parts for several tilt-to-amplitude ratios $f/u$ in the strong-modulation regime $u\gg1$. We distinguish diagonal ($\mu=\nu$) and off-diagonal ($\mu\ne\nu$) elements and vary the wave vectors $Q_\mu, Q_\nu$, and $q$.

\subsubsection{Diagonal ISF}
At the Bragg point $q=0$, the ISF does not decay to zero but approaches the finite plateau given by \cref{eq:long_time_limit}, see \cref{fig:ISF_diagonal}~(a,d). Its magnitude reflects the Fourier amplitudes of the stationary distribution and decreases with increasing $f/u$ and with larger wave vectors.

For $q\ne0$, the ISF decays to zero at long times, signaling loss of correlations at finite wavelength. The red curve in \cref{fig:ISF_diagonal}~(a-c) for $f/u=0.0$ also decays to zero in the long-time limit, but the large potential $u=U_1/k_BT=10$ makes the relaxation so slow that it lies outside the visible time range. Below the critical tilt ($f<u$), the real part shows a two-step relaxation: near-free diffusion at short times, a trapping plateau captured by the harmonic approximation, \cref{eq:ISF_HA}, around $t\simeq\tau_{\mathrm{HA}}$, \cref{eq:tau_ha}, and a final decay set by escape at $t\simeq\tau_{\mathrm{esc}}$, \cref{eq:tau_esc}, see \cref{fig:ISF_diagonal}~(a-c). In this regime the imaginary part is negligible because the particle locally equilibrates and rarely crosses barriers.

Above the critical tilt ($f>u$), a finite drift velocity $v$ induces pronounced oscillations in both real and imaginary parts, roughly consistent with a phase factor $e^{ik v_{\mathrm{det}} t}$ at high tilts. The oscillation amplitude increases with $f/u$ and decreases with larger $Q_\mu, Q_\nu$, consistent with probing shorter spatial scales over fewer potential periods, compare \cref{fig:ISF_diagonal}~(b) vs. (c) and (e) vs. (f).

In conclusion, the diagonal ISF retains full spatio-temporal information that goes beyond the analysis of low-order moments. While the variance and diffusivity follow from the long-wavelength ($q\to0$) expansion, the ISF resolves dynamics jointly in time and at a specified length scale $\simeq 2\pi/q$. In our system, it cleanly separates intra-well relaxation (short-time decay), transient trapping (plateaus), barrier crossing (late-time decay), and drift (oscillatory phase), and shows how each depends on $q$. The plateau height quantifies the degree of localization at the probed wavelength, whereas the oscillation amplitude measures how strongly directed motion (velocity) imprints on the sampled length scale. We find that the harmonic approximation, or the deterministic limit are good approximations for the corresponding regime, however in between these we rely on the ISF.

\subsubsection{Off-diagonal ISF}
 Beyond diagonal elements, the off-diagonal ISF, $F_{\mu\nu}(q,t)$, $\mu\neq \nu$, probes correlations between modes whose wave vectors differ by reciprocal-lattice increments. By periodicity, Fourier components with wave vectors $q_\mu = q + Q_\mu$ are coupled to those at $q_\nu = q + Q_\nu$, corresponding to an umklapp-like shift by a reciprocal lattice vector.

At $t=0$, the ISF is set by equilibrium mode overlaps, \cref{eq:init_ISF}, and can be positive or negative, see \cref{fig:ISF_offdiagonal}. For $q=0$, temporal correlations factorize at long times and $F_{\mu\nu}(0,t)$ approaches the nonzero constant from \cref{eq:long_time_limit}. This constant is representing the Fourier modes of the stationary solution. Furthermore, if $\mu=0$ or $\nu=0$, $F_{\mu\nu}(0,t)$ remains at its initial value for all $t$ (not shown). For $0\neq \mu\neq\nu\neq 0$, the off-diagonal ISF is essentially flat below the critical tilt, while above it exhibits transient oscillations with a phase set by the mean drift.

For $q\neq 0$, the ISF decays to zero at long times, indicating loss of inter-mode correlations; the largest deviations occur below the critical tilt, while above it oscillations appear analogous to the diagonal case. Simulations are consistent with these predictions.

\section{Conclusion}\label{sec:conclusion}
We have studied the dynamics of an overdamped Brownian particle in a tilted cosine potential and used the intermediate scattering function (ISF) to quantify spatio-temporal correlations. By exploiting periodicity through Bloch's theorem and employing a spectral formulation of the Fokker-Planck operator, we derived an analytical representation of a generalized ISF expressed in terms of the operator's eigenmodes, and we solved the associated eigenvalue problem numerically. The generalized ISF captures both diagonal and off-diagonal mode correlations, and it reduces to the standard ISF when the two wave vectors coincide.

We extend the work of Reimann \emph{et al.}~\cite{Reimann_2001,Reimann_2002} by deriving the full spatio-temporal dynamics and the complete time dependence of the variance and higher-order moments. Using time-dependent perturbation theory, we obtained low-order moments, including the mean velocity, variance, time-dependent diffusivity, skewness, and the non-Gaussian parameter. Brownian-dynamics simulations validate these predictions. The harmonic approximation provides a description of intra-well dynamics and short-time behavior, while the deterministic limit offers insight in the strong-tilt regime.
Our perturbative framework assumes the existence of a stationary state and therefore does not directly apply to explicitly time-dependent potentials. Such cases generally require numerical solutions or alternative analytical approaches, for example field-theoretic descriptions based on the Dean–Kawasaki equation~\cite{Abbott_2019,Juniper_2017,Illien_2025}.

Our analysis emphasizes the central role of the tilt-to-amplitude ratio $f/u$ in the low-temperature or high-modulation regime $u\gg1$. In the locked state with $f<u$, the ISF displays a two-step relaxation characterized by a trapping plateau around $t\gtrsim\tau_{\mathrm{HA}}$ and a subsequent decay around $t\simeq\tau_{\mathrm{esc}}$, fully consistent with intra-well equilibration followed by barrier hopping. At the Bragg point $q=0$, the ISF approaches a finite long-time plateau, whereas for $q\neq0$ it decays to zero. In this regime, the mean drift is strongly suppressed. In the running state with $f>u$, a finite drift gives rise to pronounced oscillations in the real and imaginary parts of the ISF, with phases set by the probed wave vectors; the oscillation amplitude grows with $f/u$ for diagonal elements and is reduced for off-diagonal ones. Near the critical tilt $f\approx u$, the long-time diffusivity exhibits giant diffusion, whose peak height increases and width narrows as $u$ increases.

Off-diagonal ISF components quantify inter-mode coupling and admit an umklapp-like interpretation in which correlations transfer between modes that differ by reciprocal-lattice vectors, so that 'momentum' is effectively conserved modulo a lattice vector. For $q=0$, these correlations factorize at long times and approach nonzero constants, while for $q\neq0$ they decay to zero; in the running regime, transient oscillations appear analogous to those in the diagonal case. The behavior of the higher moments reinforces this picture: in the locked state the mean velocity is suppressed, while in the running state higher moments exhibit oscillations driven by the interplay of finite drift and periodic forcing. The skewness and the non-Gaussian parameter peak slightly below the bifurcation and approach zero in both the short- and long-time limits, indicating asymptotically Gaussian behavior.

Taken together, the generalized ISF and its moment expansions provide a comprehensive and consistent description of Brownian motion in tilted periodic potentials. 
The diagonal ISF retains spatio-temporal information beyond low-order moments: while the variance and diffusivity arise from the long-wavelength ($q\to0$) limit, the ISF resolves dynamics jointly in time and at length scale $\simeq 2\pi/q$. It separates intra-well relaxation, trapping plateaus, barrier crossing, and drift via $q$-dependent decay and oscillatory phases. Plateau heights quantify localization, and oscillation amplitude quantifies directed motion. The harmonic and deterministic limits describe the respective regimes, but in between we rely on the ISF. In addition, off-diagonal ISFs expose inter-mode coupling and the ISF is directly measurable and contains unique features, such as the long-time plateau at $q=0$, which can be valuable in experimental realizations.

Within periodic settings, natural extensions include other periodic landscapes, such as asymmetric or multi-harmonic potentials, for which the same Bloch-based framework can be applied with minor modifications. By contrast, extensions to disordered or spatially inhomogeneous environments, to colored noise or weak inertia, and to interacting many-particle systems would require substantial reformulation of the present approach and are therefore speculative at this stage. 
Finally, the time-dependent solutions developed here could serve as building blocks for compact, fully analytical real-space representations, potentially reducing numerical cost in future applications.

\section*{Acknowledgements}
We thank C. Reiter for his work on the real-space coefficients (\cref{sec:real_space_coefficients}) and acknowledge his assistance with the derivations in \cref{sec:Model} and his help with the results in \cref{sec:results_ISF}.

We acknowledge the use of AI (ChatGPT, Grammarly, LanguageTool, Perplexity) for its assistance with grammar checking, translations, and text enhancement.
 
This research was funded in part by the Austrian Science Fund
(FWF) 10.55776/P35580. 

\section*{Data availability}
The data that supports the findings of this article are openly available \cite{rusch_2026_repository}.

\appendix

\section{Symmetry of the spectrum}\label{eq:spectrum}

We consider the generally non-Hermitian operator $\mathcal{L}_q$, \cref{eq:operatorLk}, which satisfies the symmetry 
\begin{align}
  \mathcal{L}_{-q} = \mathcal{L}_q^{*},
\end{align}
in particular, for $q=0$ the operator is real-valued. Taking the complex conjugate of the right-eigenvalue problem and using the symmetry above, we obtain
\begin{align}
  \mathcal{L}_{-q} u^R_{nq}(x)^* = - \lambda_{nq}^* u^R_{nq}(x)^*.
\end{align}
Thus $ u^R_{nq}(x)^*$ is a right eigenfunction of $\mathcal{L}_{-q}$ with eigenvalue $\lambda_{nq}^*$. We use the labeling convention, such that the imaginary part of the eigenvalues is continuous and find
\begin{align}
  \lambda_{nq} &= \lambda_{-n,-q}^*, \qquad 
  u_{nq}^{R/L}(x) = u_{-n,-q}^{R/L}(x)^*\nonumber
\end{align}
where the relation for the left eigenfunction is derived analogously. Because of the Bloch form of the eigenfunctions we immediately find the same relation for the eigenfunctions
\begin{align}
    \psi_{nq}^{R/L}(x)
=\psi_{-n,-q}^{R/L}(x)^*.
\end{align}
Next, we see that the operator, \cref{eq:operatorLk},
\begin{align}
  \delta\mathcal{L}_q^{1} \propto \mathrm{i} q,
\end{align}
is linear in $q$. Using the left and right eigenfunctions at $q=0$ and the above labeling, the matrix elements obey the conjugation relations
\begin{align}
  \big\langle l_{m0} \big| \delta\mathcal{L}_q^{1} \big| r_{n0} \big\rangle
  &= -\Big( \big\langle l_{-m0} \big| \delta\mathcal{L}_q^{1} \big| r_{-n0} \big\rangle \Big)^*.\nonumber
\end{align}
Consequently, products appearing in cumulant expansions, e.g.~in \cref{eq:msd}, group into complex-conjugate pairs
\begin{align}
  &\big\langle l_{00} \big| \delta\mathcal{L}_q^{1} \big| r_{0m} \big\rangle 
  \big\langle l_{m0} \big| \delta\mathcal{L}_q^{1} \big| r_{00} \big\rangle
  \\
  &= 
  \Big(
  \big\langle l_{00} \big| \delta\mathcal{L}_q^{1} \big| r_{0,-m} \big\rangle 
  \big\langle l_{-m0} \big| \delta\mathcal{L}_q^{1} \big| r_{00} \big\rangle
  \Big)^*,\nonumber
\end{align}
and therefore the cumulants are real-valued, similar to Ref.~\cite{Rusch_2024}.

\section{Real-space representation of Fourier coefficients} \label{sec:real_space_coefficients}
In this section, we derive an explicit real-space expression for the  Fourier coefficients $\langle \mu | r_{00} \rangle$. We start with the known stationary solution ~\cite{Xiao_2015,stratonovich_1967_topics}
\begin{align}
    u_{00}^R(x)= p^\text{st}(x)=\frac{ I_-(x)}{\int_0^L \diff y I_-(y) },
\end{align}
and rewrite the auxiliary function, \cref{eq:aux}
\begin{align}
I_-(x) 
&= e^{-u \cos(Q_1 x)} \int_0^L \diff y  e^{u \cos[Q_1(x + y)]} e^{- Q_1 f y},
\end{align}
using the Jacobi-Anger-Expansion
\begin{align}
    e^{u \cos[Q_1(x + y)]} &= \sum_{\mu=-\infty}^\infty I_\mu(u) e^{i \mu Q_1 x} e^{i \mu Q_1 y},\nonumber \\
        e^{-u \cos(Q_1 x)} &= \sum_{\mu=-\infty}^{\infty} (-1)^\mu I_\mu(u)  e^{i \mu Q_1 x},
\end{align}
where $I_n(\cdot)$ is the modified Bessel function of the first kind of integer order $n$. 
Inserting the Jacobi-Anger expansion and integrating the remaining exponential yields
\begin{align}
I_-(x) 
&= \sum_{\mu=-\infty}^{\infty} (-1)^\mu I_\mu(u)  e^{i \mu Q_1 x} \sum_{\nu=-\infty}^\infty I_\nu(u)  e^{i \nu Q_1 x} B_\nu \nonumber\\
&= \sum_{\mu=-\infty}^{\infty} 
\left( \sum_{\nu=-\infty}^{\infty}I_{\mu - \nu}(-u)  I_\nu(u)B_\nu \right) e^{i \mu Q_1 x},
\end{align}

where $\mu+\nu \to \mu$ and $(-1)^{\mu } I_{\mu }(u)= I_{\mu }(-u)$ was used in the last step to simplify and the coefficient~\cite{NIST:DLMF}
\begin{align}
    B_\nu= \dfrac{ \exp[{(i \nu - f) Q_1 L}]-1}{(i \nu - f) Q_1},
\end{align}
is introduced.
For the normalization we use that 
\begin{align}
\int_0^L\diff y \, e^{i(\nu - \mu) Q_1 y} =  L  \delta_{\nu\mu} ,
\end{align}
and therefore
\begin{align}
\int_0^L \diff y \, I_-(y)
&= L \sum_{\sigma=-\infty}^{\infty} I_\sigma(u)^2  B_\sigma.
\end{align}
The stationary solution has the Fourier representation
\begin{align}
    p^{\text{st}}(x) = \sum_{\mu=-\infty}^{\infty} \frac{e^{i \mu Q_1 x}}{\sqrt{L}} \langle \mu | r_{00} \rangle,
\end{align}
with Fourier coefficients
\begin{align}
    \langle \mu | r_{00} \rangle= \frac{1}{\sqrt{L}} \frac{\sum_{\nu=-\infty}^\infty  I_{\mu - \nu}(-u)  I_\nu(u)B_\nu }{\sum_{\sigma=-\infty}^{\infty} I_\sigma(u)^2  B_\sigma}.
\end{align}

\section{Cumulants and moments} \label{sec:cumulants}

For the case of the linear operator, $\delta \mathcal L^1_q \propto q$, the reduced cumulants can be directly read off by comparing \cref{eq:cumulant_expansion,eq:explicit_ISF_cumulant}. For clarity we state the reduced cumulants explicitly

\begin{widetext}
    
\begin{align} \label{eq:explicit_cumulants}
- i q   \tilde{\kappa}_1[\Delta x(t)] &= 	t \langle l_{00} |   \delta \mathcal{L}_{q}^1   r_{00}\rangle\\
- \frac{q^2}{2}  \tilde{\kappa}_2[\Delta x(t)]   &=  
	\sum_{n \neq 0} \frac{e^{-\lambda_{n0} t} + \lambda_{n0} t -1}{\lambda_{n0}^2}  
	\langle l_{00} |\delta \mathcal{L}_{q}^1  r_{n0} \rangle \langle l_{n0} |  \delta\mathcal{L}_{q}^1   r_{00} \rangle  \nonumber \\
+ \frac{i q^3}{3!}  \tilde{\kappa}_3[\Delta x(t)] &=  
	\sum_{n \neq 0}   \frac{\lambda_{n0}  t+e^{-\lambda_{n0}  t} (\lambda_{n0}  t+2)-2}{\lambda_{n0} ^3} \langle l_{00} |\delta \mathcal{L}_{q}^1  r_{n0} \rangle \langle l_{n0} |  \delta\mathcal{L}_{q}^1   r_{00} \rangle  (\langle l_{n0} | \delta \mathcal{L}_{q}^1  r_{n0} \rangle  -\langle l_{00} | \delta \mathcal{L}_{q}^1  r_{00} \rangle 	) \nonumber \\
	&+
	\sum_{n \neq 0} \sum_{m \neq 0,m \neq n } \left(\frac{e^{-\lambda_{n0} t}+\lambda_{n0}  t-1}{\lambda_{n0} ^2 (\lambda_{m0} -\lambda_{n0} )} + \frac{e^{-\lambda_{n0} t}+\lambda_{n0}  t-1}{\lambda_{m0} ^2 (\lambda_{n0} -\lambda_{m0} )}\right)  \langle l_{00} | \delta \mathcal{L}_{q}^1   r_{n0} \rangle \langle l_{n0} | \delta \mathcal{L}_{q}^1    r_{m0} \rangle \langle l_{m0} | \delta \mathcal{L}_{q}^1  r_{00} \rangle,\nonumber .
\end{align} 

\end{widetext}

For the fourth cumulant, we first compute the corresponding reduced moments. The ISF, \cref{eq:reduced_ISF}, expanded in its moments up to the fourth order is given by

\begin{widetext}
    
\begin{align}\label{eq:moments}
\tilde{F}(q, t)=&\; 1
+t\left\langle l_{00} | \delta \mathcal{L}_q^{1} r_{00}\right\rangle
+\sum_n \frac{e^{-\lambda_{n0} t}+\lambda_{n0} t-1}{\lambda_{n0}^2}
\left\langle l_{00} | \delta \mathcal{L}_q^{1} r_{n0}\right\rangle
\left\langle l_{n0} | \delta \mathcal{L}_q^{1} r_{00}\right\rangle \\
& +\sum_n \sum_m
\left(
\frac{e^{-\lambda_{n0} t}+\lambda_{n0} t-1}{\lambda_{n0}^2(\lambda_{m0}-\lambda_{n0})}
+\frac{e^{-\lambda_{m0} t}+\lambda_{m0} t-1}{\lambda_{m0}^2(\lambda_{n0}-\lambda_{m0})}
\right)
\left\langle l_{00} | \delta \mathcal{L}_q^{1} r_{n0}\right\rangle
\left\langle l_{n0} | \delta \mathcal{L}_q^{1} r_{m0}\right\rangle
\left\langle l_{m0} | \delta \mathcal{L}_q^{1} r_{00}\right\rangle \nonumber \\
& +\sum_n \sum_m \sum_p
\left(
\frac{e^{-\lambda_{n0} t}+\lambda_{n0} t-1}
{\lambda_{n0}^2(\lambda_{n0}-\lambda_{m0})(\lambda_{n0}-\lambda_{p0})}
+\frac{e^{-\lambda_{m0} t}+\lambda_{m0} t-1}
{\lambda_{m0}^2(\lambda_{m0}-\lambda_{n0})(\lambda_{m0}-\lambda_{p0})}
+\frac{e^{-\lambda_{p0} t}+\lambda_{p0} t-1}
{\lambda_{p0}^2(\lambda_{p0}-\lambda_{m0})(\lambda_{p0}-\lambda_{n0})}
\right) \nonumber \\
& \qquad \times
\left\langle l_{00} | \delta \mathcal{L}_q^{1} r_{n0}\right\rangle
\left\langle l_{n0} | \delta \mathcal{L}_q^{1} r_{m0}\right\rangle
\left\langle l_{m0} | \delta \mathcal{L}_q^{1} r_{p0}\right\rangle
\left\langle l_{p0} | \delta \mathcal{L}_q^{1} r_{00}\right\rangle
+O(q^5).
\nonumber
\end{align}

\end{widetext}

where the sums over $n, m, p$  formally include all bands. This can produce apparent zero denominators when an index hits the lowest band or when two indices coincide. In each such instance, however, the corresponding numerators vanish as well, so the singularities are removable.

The way we use to eliminate zero divisors is to separate out these special index configurations before performing the integrals and treat them case by case. So, we partition the sums so that terms with potential $0/0$ structure are isolated and evaluated via their limiting forms, as explicitly shown in \cref{eq:explicit_cumulants,eq:explicit_ISF_cumulant}. In \cref{eq:moments}, however, this procedure gets lengthy for the fourth moment, and we refer to the appendix of Ref.~\cite{Rusch_2024}, where the explicit form is given with the difference that we use the operator $ \delta \mathcal{L}_{q}^1 $ and expand the \emph{reduced} ISF.

A further subtle point arises for the pure cosine potential at $f=0$. Except for the ground state, all eigenvalues $\lambda_{0n}$ become twofold degenerate. This degeneracy generates additional zero denominators that must be handled carefully. Our treatment follows the procedure detailed in Ref.~\cite{Rusch_2025}.

\section{Long-time Mobility} \label{sec:GER}
To compute the mobility, \cref{eq:mobility_def}, we differentiate the velocity, \cref{eq:velocity}, with respect to the tilt $f$. Since both the operator $\delta\mathcal{L}_q^{1}$ and the right eigenstates depend on $f$, we obtain
\begin{align}
    \frac{\partial v}{\partial f}
    = \frac{i}{q}\Big[
        \big\langle l_{00} \big| \tfrac{\partial \delta\mathcal{L}_q^{1}}{\partial f} \big| r_{00} \big\rangle
        + \big\langle l_{00} \big| \delta\mathcal{L}_q^{1} \big| \tfrac{\partial r_{00}}{\partial f} \big\rangle
    \Big].
    \label{eq:dvdf_three_terms}
\end{align}
We already use that $\langle l_{00}|$ does not depend on $f$ as evident in the real-space representation $\langle x |l_{00}\rangle = 1$. For the first term we compute
\begin{align}
    \frac{\partial \delta\mathcal{L}_q^{1}}{\partial f} = - D Q_1\, i q,
\end{align}
which immediately yields
\begin{align}
    \frac{i}{q}\,
    \big\langle l_{00} \big| \tfrac{\partial \delta\mathcal{L}_q^{1}}{\partial f} \big| r_{00} \big\rangle
    = D Q_1 \big\langle l_{00} \big| r_{00} \big\rangle
    = D Q_1.
\end{align}
For the second term, we start from the right zero mode,
\begin{align}
    \mathcal{L}_0 \big| r_{00} \big\rangle = 0,
\end{align}
and differentiate with respect to $f$ to obtain
\begin{align}
    \mathcal{L}_0 \left| \frac{\partial r_{00}}{\partial f} \right\rangle
    = - \frac{\partial \mathcal{L}_0}{\partial f} \big| r_{00} \big\rangle.
    \label{eq:inhomogeneous}
\end{align}
Expanding in the complete bi-orthonormal basis,
\begin{align}
    \left| \frac{\partial r_{00}}{\partial f} \right\rangle
    = \sum_{n} c_n \big| r_{n0} \big\rangle,
\end{align}
and using $\mathcal{L}_0 \big| r_{n0} \big\rangle = - \lambda_{n0} \big| r_{n0} \big\rangle$, projection onto $\langle l_{m0}|$ gives
\begin{align}
    c_m = \frac{
        \big\langle l_{m0} \big| \tfrac{\partial \mathcal{L}_0}{\partial f} \big| r_{00} \big\rangle
    }{\lambda_{m0}},
    \qquad m \neq 0,
\end{align}
using $\langle l_{m0} | r_{n0} \rangle = \delta_{nm}$. Hence,
\begin{align}
    \left| \frac{\partial r_{00}}{\partial f} \right\rangle
    = \sum_{n \neq 0}
    \frac{
        \big\langle l_{n0} \big| \tfrac{\partial \mathcal{L}_0}{\partial f} \big| r_{00} \big\rangle
    }{\lambda_{n0}}
    \big| r_{n0} \big\rangle
    + c_0 \big| r_{00} \big\rangle.
    \label{eq:dr00df}
\end{align}
From the normalization condition
\begin{align}
  \big\langle l_{00} \big| r_{00} \big\rangle = 1
  \quad \Rightarrow \quad
  \frac{d}{df} \big\langle l_{00} \big| r_{00} \big\rangle = 0,
\end{align}
and differentiating, we find
\begin{align}
\frac{\partial}{\partial f}\big\langle l_{00} \big| r_{00} \big\rangle
= 0 + \big\langle l_{00} \big| \tfrac{\partial r_{00}}{\partial f} \big\rangle
= 0.
\end{align}
Comparing with \cref{eq:dr00df}, we conclude that
\begin{align}
    c_0 = 0.
\end{align}
Using the Fourier basis, \cref{eq:real_space_rep}, the matrix elements of $\partial \mathcal{L}_0 / \partial f$ (cf. \cref{eq:matrix_elements}) read
\begin{align}
    \label{eq:matrix_elements_derivative}
    \big\langle \mu \big|
    \frac{\partial \mathcal{L}_0}{\partial f}
    \big| \nu \big\rangle
    = - D Q_1^{2}\, i \nu \,\delta_{\mu \nu}.
\end{align}
Expanding in this basis, the remaining matrix element becomes
\begin{align}
    \big\langle l_{n0} \big| \tfrac{\partial \mathcal{L}_0}{\partial f} \big| r_{00} \big\rangle
    = - D Q_1^{2} \sum_{\nu} i \nu \,\big\langle l_{n0} \big| \nu \big\rangle \big\langle \nu \big| r_{00} \big\rangle.
    \label{eq:mel_fourier}
\end{align}
The effective mobility is then
\begin{align}
    k_B T\, \mu_{\infty}
    = D
    + \frac{i}{q Q_1}\sum_{n \neq 0}
    \frac{
        \big\langle l_{n0} \big| \tfrac{\partial \mathcal{L}_0}{\partial f} \big| r_{00} \big\rangle\,
        \big\langle l_{00} \big| \delta\mathcal{L}_q^{1} \big| r_{n0} \big\rangle
    }{\lambda_{n0}}.
\end{align}
We now specialize to the equilibrium case, $f=0$, where this equation simplifies using
\begin{align} \label{eq:PL_eq}
\frac{1}{Q_1}   \left.   \frac{\partial \mathcal{L}_0}{\partial f}  \right|_{f=0} \langle x| r_{00} \big\rangle
   = -D\,  \partial_x \langle x| r_{00} \big\rangle.
\end{align}
An analogous calculation for $\delta\mathcal{L}_q^{1}$ together with $\langle x|r_{00}\big\rangle=p^{\text{eq}}(x)$  and  $
\partial_x p^{\text{eq}}(x) = -U'(x)/(k_B T) \, p^{\text{eq}}(x)$ gives
\begin{align} \label{eq:Lq1_eq}
   -\frac{1}{i q}\,\left.  \delta\mathcal{L}_q^{1} \right|_{f=0} \langle x | r_{00} \rangle 
   &= \big(D Q_1 u \sin(Q_1 x) - 2D\,  \partial_x \big) \langle x | r_{00} \rangle \nonumber \\
   &= -  D\, \partial_x \langle x | r_{00} \rangle.
\end{align}
Since both expressions evaluate to $-  D\, \partial_x \langle x | r_{00} \rangle$, the corresponding matrix elements satisfy
\begin{align}
\label{eq:mu-eff-again}
 \frac{1}{Q_1} \left.\, \big\langle l_{n0} \big| \tfrac{\partial \mathcal{L}_0}{\partial f} \big| r_{00} \big\rangle \right|_{f=0}  = 
 -\frac{1}{i q}\, \left. \big\langle l_{n0} \big| \delta\mathcal{L}_q^{1} \big| r_{00} \big\rangle \right|_{f=0}.
\end{align}
Substituting into the expression of $\mu_\infty$, \cref{eq:mobility}, and comparing with \cref{eq:diffusivity_explicitely_longtime}, this identity yields the equilibrium Einstein relation 
\begin{align}
  k_B T\, \mu_\infty = D_\infty.
\end{align}



%

\end{document}